# Pressure-induced superconductivity in itinerant antiferromagnet CrB$_2$

Cuiying Pei[1#], Pengtao Yang[2,3#], Chunsheng Gong[4#], Qi Wang[1,5], Yi Zhao[1], Lingling Gao[1], Keyu Chen[2,3], Qiangwei Yin[4], Shangjie Tian[4], Changhua Li[1], Weizheng Cao[1], Hechang Lei[4*], Jinguang Cheng[2,3*], and Yanpeng Qi[1*]

1. School of Physical Science and Technology, ShanghaiTech University, Shanghai 201210, China
2. Beijing National Laboratory for Condensed Matter Physics and Institute of Physics, Chinese Academy of Sciences, Beijing 100190, China
3. School of Physical Sciences, University of Chinese Academy of Sciences, Beijing 100190, China
4. Department of Physics and Beijing Key Laboratory of Opto-electronic Functional Materials & Micro-nano Devices, Renmin University of China, Beijing 100872, China
5. ShanghaiTech Laboratory for Topological Physics, ShanghaiTech University, Shanghai 201210, China

# These authors contributed to this work equally.
* Correspondence should be addressed to Y.P.Q. (qiyp@shanghaitech.edu.cn) or J-G. C. (jgcheng@iphy.ac.cn) or H.C.L. (hlei@ruc.edu.cn)

**ABSTRACT:**

**The recent discovery of superconductivity up to 32 K in the pressurized MoB$_2$ revives the interests in exploring novel superconductors in transition-metal diborides isostructural to MgB$_2$. Although the Mo 4$d$-electrons participate in Cooper pairing, the electron-phonon coupling remains the dominant mechanism for the emergence of superconductivity in MoB$_2$. To explore possible superconductivity driven by unconventional pairing mechanism, we turn our attention to an itinerant antiferromagnet CrB$_2$. Here we report on the discovery of superconductivity up to 7 K in CrB$_2$ via the application of external pressure. Superconductivity is observed after the antiferromagnetic transition at $T_N$ ~ 88 K under ambient pressure is completely suppressed. Then, the superconducting $T_c$ increases monotonically with pressure and this evolution takes place without a structural transition in the sample. Since the proximity of superconductivity to an antiferromagnetic order, the quantum criticality and unconventional superconductivity may exist in CrB$_2$. Current study would promote further studies and explorations of novel superconductors in other transition-metal diborides.**

# INTRODUCTION

Now it is generally accepted that superconductivity occurs via two microscopic mechanisms. One is the celebrated Bardeen-Cooper-Schrieffer (BCS) mechanism based on the electron-phonon coupling that can be applied to the large family of so-called conventional superconductors, among which $MgB_2$ holds the record of the highest $T_c$ at ambient pressure reported so far. Thus, the $T_c \approx 39$ K of $MgB_2$ has been considered to approach the McMillan limit[1]. The others beyond the BCS theory are usually named as unconventional superconductors, among which the well-known examples include the cuprates and iron-based high-$T_c$ superconductors with the $T_c$ values well above the McMillan limit[2, 3]. Although the microscopic mechanism for these unconventional superconducting families remains elusive so far, one of the common features is that the superconductivity emerges in the vicinity of long-range antiferromagnetically ordered state [4-7]. The close proximity of superconductivity to a magnetic instability suggests that the critical spin fluctuations should play a crucial role on the formation of Cooper pairs. As such, to realize an antiferromagnetic quantum critical point (QCP) has been considered as an effective approach for exploring novel unconventional superconductors. Despite of the prevailing consensus on the magnetism-mediate mechanism, recent studies on several unconventional superconductors have also indicated the importance of electron-phonon coupling[8, 9]. In this regard, it would be interesting to explore some candidate superconducting materials that can potentially combine cooperatively the phonon- and spin-mediate mechanisms.

Keeping this idea in mind, we recently initiated a high-pressure study on the transition-metal diborides, which are isostructural to $MgB_2$ but are expected to have the spin degree of freedom. We surprisingly discovered that the application of high pressure drives the paramagnetic $MoB_2$ to a superconductor with $T_c$ as high as 32 K, which is close to that of $MgB_2$ and the highest among the transition-metal diborides even though the Mo-4$d$ electrons dominate the density of states near Fermi level. Detailed analyses revealed that the $d$-electrons and phonon modes of transition metal Mo atoms play utterly important roles in the emergence of superconductivity in $MoB_2$ [10]. This discovery of superconductivity in $MoB_2$ has rekindled enthusiasm on explorations of more superconductors among the transition-metal diborides. Recently, Lim et al. reported the discovery of superconductivity in $WB_2$ with the optimal $T_c \sim 17$ K at 90 GPa [11]. Although the 4$d$/5$d$-electrons participate in the formation of Cooper pairs in these two compounds, the influences of spin degree of freedom are not evident due to the absence of long-rage magnetic order at ambient conditions.

To this end, we turn our attention to a 3$d$ transition-metal diborides, $CrB_2$, which exhibits an itinerant-electron antiferromagnetic (AF) transition at $T_N \approx 88$ K at ambient pressure[12]. $CrB_2$ has been known for a long time and it adopts the $AlB_2$-type hexagonal crystal structure (space group: *P6/mmm*), isostructural to the well-known $MgB_2$[13]. Although some high-pressure study at very limited pressure range has been performed on this compound [14], none of the following questions have been addressed so far: (i) whether the AF order can be suppressed completely by higher pressure? (ii) if so, can the superconductivity emerge near the AF QCP? (iii) what is the role of spin and phonon in mediating Cooper pairs?

To address these issues, here we perform a comprehensive high-pressure study on the $CrB_2$ single crystal over a wide pressure range. We find that the AF order of $CrB_2$ can indeed be suppressed by pressure ~ 4 GPa and then bulk superconductivity appears at higher pressures. The superconducting $T_c$ increases monotonically with pressure and reaches a maximum and unsaturated $T_c$ of 7 K at ~ 100 GPa without showing structural transition. In addition to CrAs, $CrB_2$ becomes the second Cr-based superconductor at high pressures.

**EXPERIMENTAL SECTION**

Single crystals of $CrB_2$ were grown out of Al flux. Cr (99.5 %), B (99.9 %) and Al (99.99 %) with a molar ratio of Cr : B : Al = 1 : 2.5 : 73.3 were put into an alumina crucible. The mixture was heated up to 1773 K in a high-purity argon atmosphere. Then it was cooled down to 1173 K. The $CrB_2$ single crystals were separated from the Al flux using sodium hydroxide solution. The single crystal X-ray diffraction (XRD) pattern was performed using a Bruker D8 X-ray diffractometer with Cu $K_\alpha$ radiation ($\lambda$ = 0.15418 nm) at room temperature.

The high-pressure resistivity was measured with the standard four-probe method in a palm-type cubic anvil cell (CAC) apparatus at pressures < 12 GPa[15]. Glycerol was used as the pressure transmitting medium. The pressure values were estimated from the pressure-load calibration curve determined at low temperatures by monitoring the superconducting transition of lead. It should be noted that the pressure values inside the CAC exhibit slight variations upon cooling, which has been well characterized in our previous work[15].

Resistivity measurements under higher pressures (>12 GPa) were performed in a nonmagnetic diamond anvil cell (DAC). A cubic BN/epoxy mixture layer was inserted between BeCu gaskets and electrical leads. Electrical resistivity was measured using the dc current in van der Pauw technique in Physical Property Measurement System (Quantum Design). Pressure was measured at room temperature using the ruby scale by measuring the luminescence from small chips of ruby placed near the sample.

In-situ high pressure XRD measurements were performed at the beamline 15U at Shanghai Synchrotron Radiation Facility ($\lambda$ = 0.6199 Å). Symmetric DAC with anvil culet size of 200 μm and Re gasket were used. Mineral oil was used as pressure transmitting medium and pressure was determined by the ruby luminescence method[16]. $CeO_2$ was used to calibrate the sample-detector distance and the orientation parameters of the detector. The two-dimensional diffraction images were analyzed using the FIT2D program[17]. Rietveld refinements on crystal structures under high pressure were performed by General Structure Analysis System (GSAS) and graphical user interface EXPGUI package[18, 19].

**RESULTS AND DISCUSSION**

At ambient pressure, the resistivity $\rho(T)$ of $CrB_2$ shows a metallic behavior and displays a pronounced kink anomaly at $T_N \approx 88.5$ K corresponding to the AF transition temperature. The AF order is further confirmed by magnetic measurements and is consistent with previous studies[20]. To study the effect of pressure on the AF transition of $CrB_2$, we have measured the temperature-dependent resistivity $\rho(T)$ under various hydrostatic pressures up to 12 GPa (Fig. 1a) by using the CAC. With increasing pressure, the magnitude of resistivity shifts down slightly in the whole temperature range, but $T_N$ manifested by the peak in $d\rho/dT$ is reduced quickly to ~ 66.2 K at 2 GPa and 42.3 K at 3.5 GPa, and almost vanishes over 4 GPa, as shown in Figs. 1b and 1c. The pressure dependence of $T_N$ shown in Fig. 1d indicates the presence of AF QCP near ~ 4 GPa. In contrary to our expectation, however, we did not observe superconductivity down to 1.4 K and up to 12 GPa, which are the temperature and pressure limit of our CAC apparatus. Whether the superconductivity will occur at lower temperatures in the studied pressure range deserves further studies.

To explore whether superconductivity can be achieved in $CrB_2$ under higher pressure, we then measured $\rho(T)$ of $CrB_2$ single crystal by using DAC in an extended pressure range. Figure 2a shows the typical $\rho(T)$ curves for pressures up to 110 GPa. When the pressure increases to 17.9 GPa, a small drop of $\rho$ is observed at the lowest measuring temperature ($T$ = 1.8 K), as shown in Fig. 2b. With further increasing pressure, the drop becomes more obvious and zero resistivity is achieved at low temperature for $P$ > 59.1 GPa, indicating the emergence of superconductivity. The superconducting $T_c$ increases monotonically with pressure and reaches about 7.3 K at $P$ = 110.4 GPa, which is the pressure limit of DAC in our present study. It is noteworthy that $T_c$ still shows an increasing trend without levelling off with pressure. The measurements on different samples of $CrB_2$ in three independent runs provide consistent and reproducible results, confirming the intrinsic superconductivity under pressure.

To gain insights into the superconducting transition, we measured $\rho(T)$ curves of $CrB_2$

under various magnetic fields at 110.4 GPa. Figure 2c demonstrates that the resistivity drop is continuously suppressed with increasing magnetic field and the superconducting transition could not be observed above 1.8 K at $\mu_0 H = 7$ T. Such behavior further confirms that the sharp decrease of $\rho(T)$ should originate from a superconducting transition. The derived upper critical field $\mu_0 H_{c2}(T)$ as a function of temperature $T$ can be well fitted by using the empirical Ginzburg-Landau formula (Figure 2d) $\mu_0 H_{c2}(T) = \mu_0 H_{c2}(0)(1 - t^2)/(1 + t^2)$, where $t = T/T_c$ is the reduced temperature with zero-field superconducting $T_c$. The obtained zero-temperature upper critical field $\mu_0 H_{c2}(0)$ of $CrB_2$ from the 90% $\rho_n$ criterion can reach 5.81 T at 110.4 GPa, which yields a Ginzburg–Landau coherence length $\xi_{GL}(0)$ of 7.54 nm.

At ambient pressure, $CrB_2$ crystallize in a typical $AlB_2$-type structure with space group *P6/mmm*. To check the structure stability under pressure, *in situ* XRD measurements on pulverized $CrB_2$ have been performed under various pressures at room temperature. Figure 3a displays the high-pressure synchrotron XRD patterns of $CrB_2$ measured at room temperature up to 97.2 GPa. All the patterns are well described by the ambient-pressure $AlB_2$ structure without showing any structural transition up to nearly 100 GPa. The representative refinements at 1 atm and selected pressures are displayed in Supplemental Information. As shown in Figure 3b, both *a*- and *c*-axis lattice constants decrease with increasing pressure, and the *c* axis shrinks significantly with the rate of ~10.2%. The large contraction of *c*-axis indicates the enhanced graphitic boron layers coupling along the *c*-axis.

From the above measurements, we can conclude that the application of high pressure first suppresses the long-range AF order around $P_c \sim 4$ GPa and then induces superconductivity with $T_c$ increasing continuously with pressure. As shown in Figure 4, a rough extrapolation of the $T_c(P)$ to zero seems to match closely to $P_c$. In addition, the low-temperature $\rho(T)$ change from Fermi liquid to non-Fermi liquid behaviors near $P_c$ and then back to Fermi-liquid-like one, implying an intimated correlation between the AF QCP and the emergence of superconductivity. Unlike the common observations that $T_c$ peaks out near the AF QCP, however, $T_c(P)$ continuously climbs up with increasing pressure without showing signs of saturation up to at least 110 GPa. This observation implies that besides the possible magnetism-mediate pairing mechanism, other pairing mechanism may also exist in $CrB_2$, which need to be clarified in the future.

Chromium (Cr) is the only element metal that shows spin-density-wave AF order above room temperature, and superconductivity has not been observed even when the AF order is suppressed by either high pressure or chemical doping[21-23]. Except for several binary Cr alloys[24-28], there have been very few Cr-containing compounds exhibiting superconductivity[29-34]. Besides CrAs, $CrB_2$ is the second Cr-based superconductor under high pressures. For CrAs, the superconductivity emerges with the maximum $T_c$ close to the AF QCP, while the optimal $T_c$ for $CrB_2$ occurs well above $P_c$. From a structural perspective, Cr-based superconductors up to now are limited to one-dimensional ($A_2Cr_3As_3$ and $ACr_3As_3$ (A = K, Rb, Cs, Na)) or three-dimensional structures (CrAs, $Pr_3Cr_{10-x}N_{11}$, $Cr_2Re_3B$). $CrB_2$ shown here has a typical layered structure, which resembles the situation in the 2D cuprates and iron-pnictides. Current study thus sheds light on the exploration of novel superconductors in the Cr and other transition metal-based systems.

## ACKNOWLEDGMENT

This work was supported by the National Key R&D Program of China (Grant No. 2018YFA0704300, 2018YFE0202600, 2018YFA0305700), the National Natural Science Foundation of China (Grant No. U1932217, 11974246, 12004252, 11822412, 11774423, 12025408, 11921004, 11834016), the Natural Science Foundation of Shanghai (Grant No. 19ZR1477300), the Science and Technology Commission of Shanghai Municipality (19JC1413900), Beijing Natural Science Foundation (Grant No. Z200005, Z190008) , the Strategic Priority Research Program of CAS (Grant No. XDB25000000, and XDB33000000) and the CAS Interdisciplinary Innovation Team. The authors thank the support from Analytical Instrumentation Center (# SPST-AIC10112914), SPST, ShanghaiTech University. The authors thank the staffs from BL15U1 at Shanghai Synchrotron Radiation Facility for assistance during data collection.

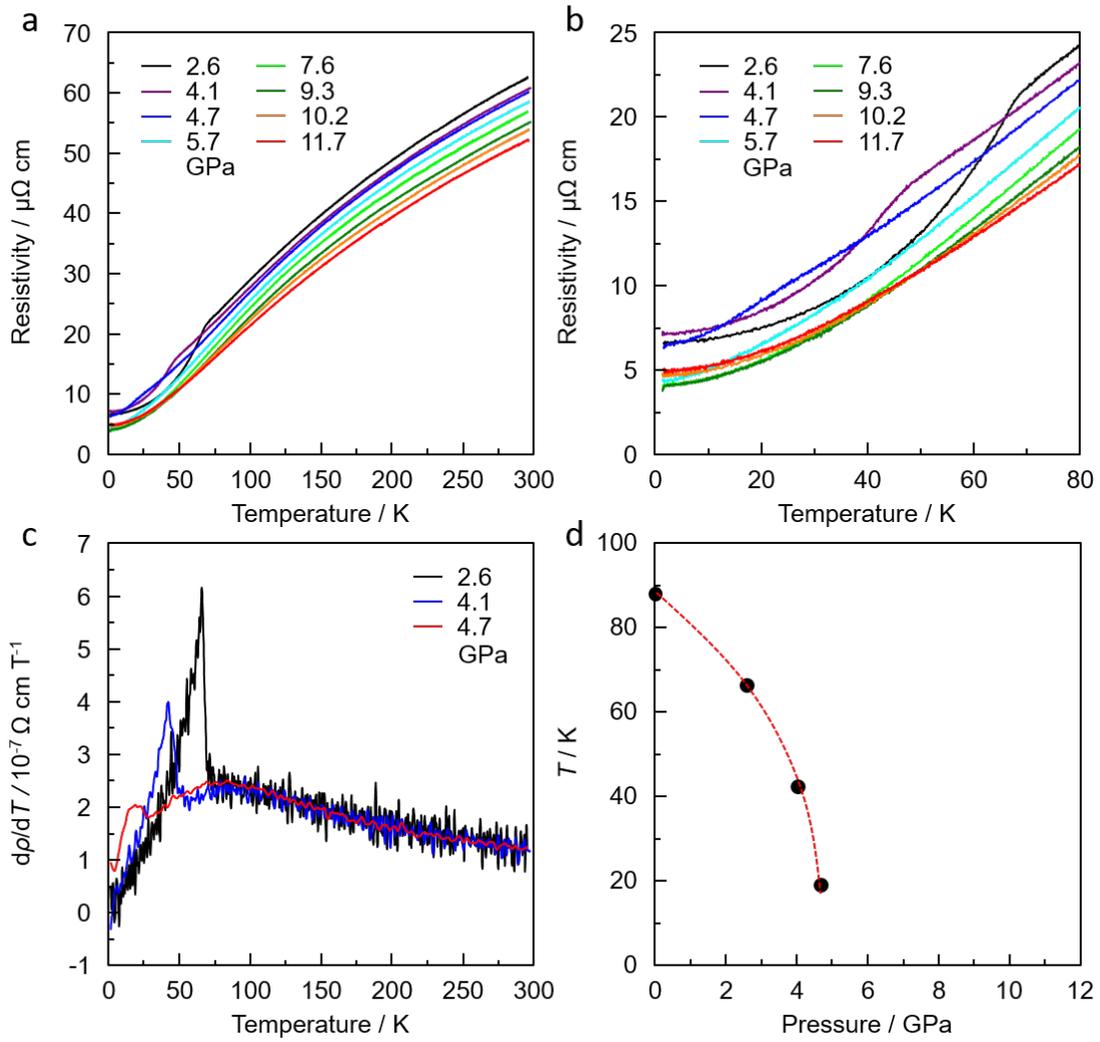

Figure 1. Temperature-dependent (a, b) resistivity $\rho(T)$ under hydrostatic pressures up to 11.7 GPa, (c) its derivative $d\rho/dT$ of $CrB_2$ under pressures of 2.6, 4.1 and 4.7 GPa; The AF transition temperature $T_N$ determined from the maximum of $d\rho/dT$ is shown in (d) as a function of pressure.

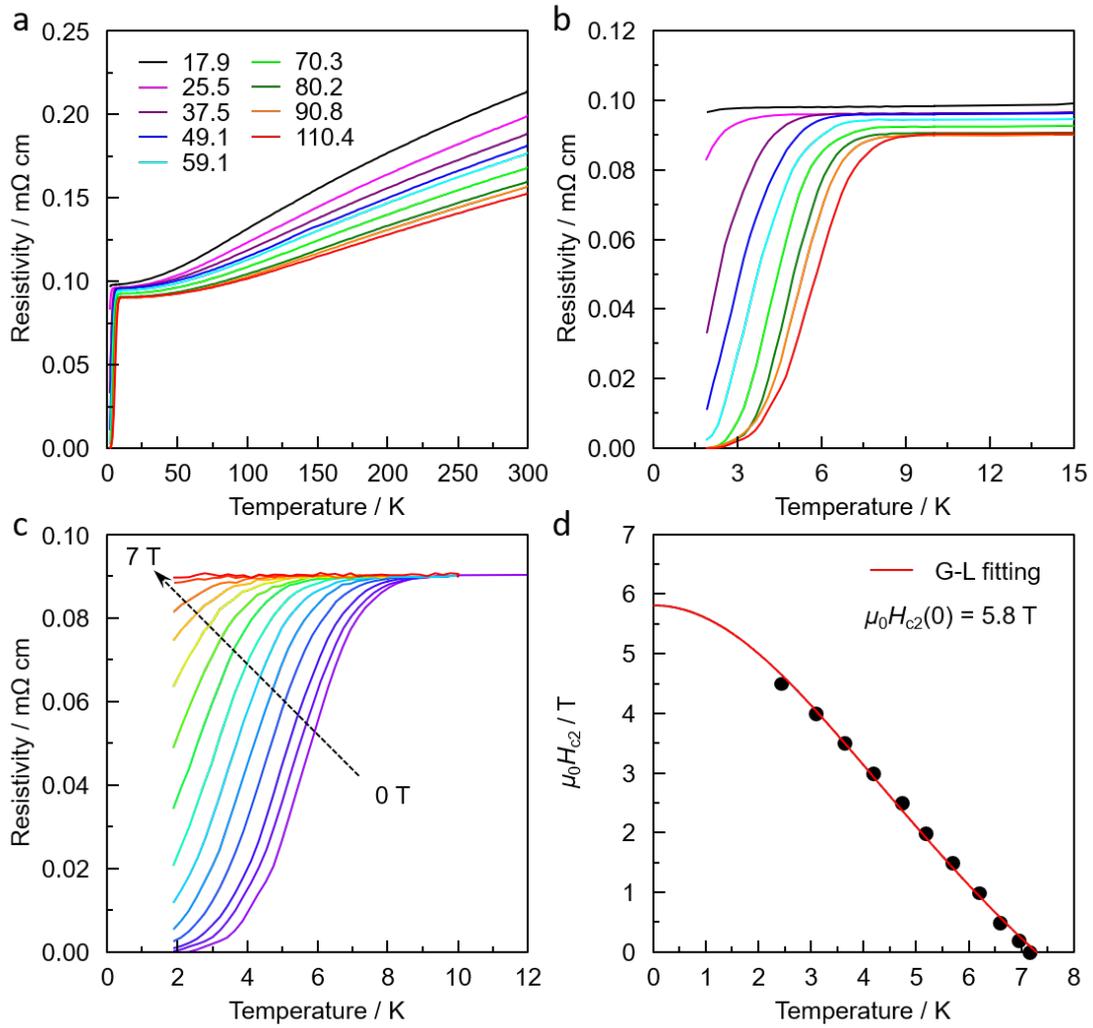

Figure 2. Transport properties of CrB$_2$ as functions of pressure. (a) Electrical resistivity $\rho(T)$ of CrB$_2$ as a function of temperature at different pressures. (b) Enlarged $\rho(T)$ curves in the vicinity of the superconducting transition. (c) $\rho(T)$ under various magnetic fields at 110.4 GPa. (d) Temperature dependence of upper critical field $\mu_0H_{c2}(T)$ at 110.4 GPa. Here, the $T_c$s are determined at the 90% of the normal state resistivity just above the onset superconducting transition temperature. The red lines represent the fits using the Ginzburg-Landau formula.

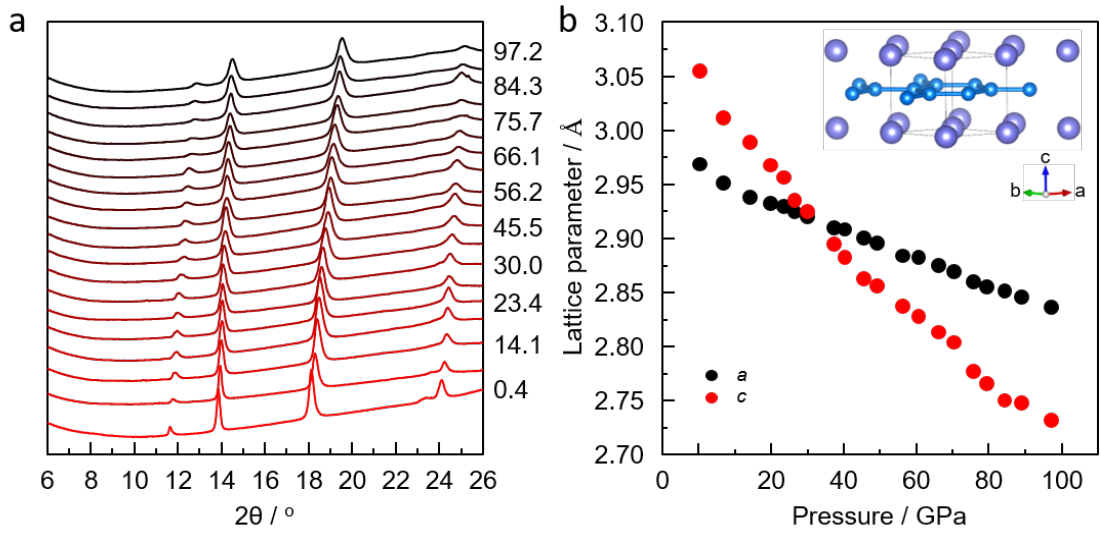

Figure 3. (a) XRD patterns of CrB$_2$ under pressure at room temperature with an x-ray wavelength of $\lambda$ = 0.6199 Å. (b) Pressure-dependence of *a*- and *c*- lattice constants for CrB$_2$. Insets: the schematic crystal structure of CrB$_2$.

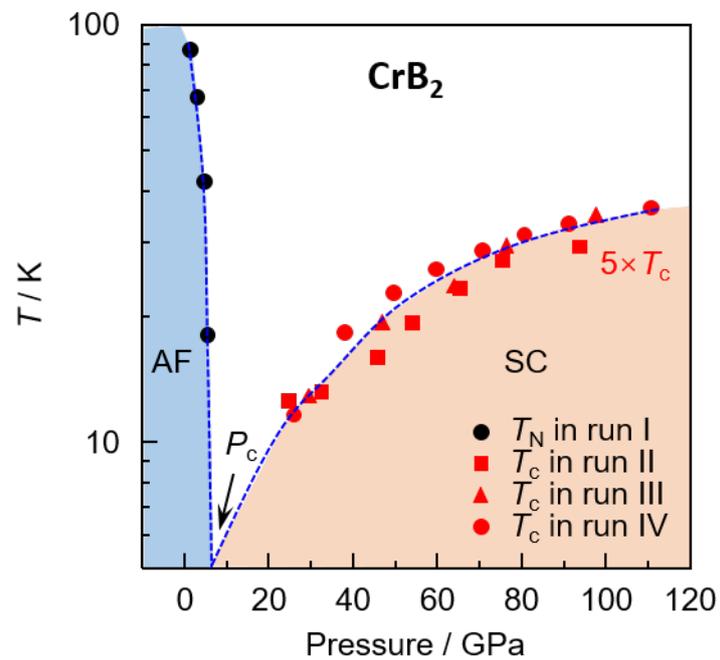

Figure 4. Temperature-pressure phase diagram of CrB$_2$.